\documentclass[aps,prd,twocolumn,nofootinbib]{revtex4-2}
\usepackage{natbib}
\usepackage{ragged2e} 

\begin{document}
\title{P and CP solution of the Strong CP puzzle}
\author{Ravi Kuchimanchi}
\email{raviparity@gmail.com}
\begin{abstract}
We use parity (P) to set $\theta_{QCD}$ to zero in the minimal left-right symmetric model with a bi-doublet Higgs, add a heavy vectorlike quark family, and obtain in a novel manner the Nelson Barr (NB) form associated so far only with spontaneous CP solution to the strong CP Puzzle. Our solution does not have the `coincidence of scales problem', that typically plagues NB models. P protects $\bar{\theta}$, if it breaks at a scale $v_R$  below the mass $M$ of the heavy quarks,  and  $\bar{\theta} \sim 10^{-9} (v_R/M)^2$ is radiatively generated, which can be acceptably small. On the other hand, if $M < v_R$, the $\bar{\theta} \sim 10^{-9}$ generated by the NB mechanism is too large, but if $\delta_{CKM}$ is obtained without the NB form, surprisingly a lower irreducible $\bar{\theta} \sim (10^{-13}~to~10^{-10}) ln( {v_R/M)}$,  testable by neutron EDM experiments is generated. No leptonic CP violation is generated (Dirac phase $\delta_{CP} = 0~or~\pi$ in PMNS matrix) which makes the minimal model testable by neutrino experiments. We also discuss some challenges in a non-minimal model that generates leptonic CP violation. Lastly but importantly, we  find with doublet rather than bi-doublet Higgses, that there is an \textit{automatic} NB solution on imposing CP (the NB form is accidental due to $SU(2)_R$),  which does not require generalized parity and needs just one mirror generation.   
\end{abstract}
\maketitle

\section{Introduction}

Before the discovery of parity violation by weak interactions, it was thought that discrete spacetime symmetries of spatial reflection or parity (P) and time reversal  (T or equivalently CP due to CPT theorem) are respected by all the forces of nature.  

After parity violation and the Standard Model (SM) were established,  an idea emerged~\cite{Weinberg_2004} that symmetries like P  (or left-right symmetry), C or other global symmetries are accidental symmetries of the strong and electromagnetic interactions, and there is no deep reason why they should be exactly conserved in nature in its higher energy theory. 

A problem with this idea is that then the Strong CP phase $\bar{\theta}$ should have been $\sim 1$ like  $\delta_{CKM}$, as there is no accidental symmetry of the SM that makes it vanish.  
However, neutron EDM experiments~\cite{PhysRevLett.124.081803} have established $\bar{\theta}\leq 10^{-10}$ radians (in its magnitude, or mod $\pi$).

It is therefore possible that either P or CP or both are symmetries of nature in a high energy theory above the SM. Since $\bar{\theta}$ is odd under $P$ as well as $CP$, it would vanish and be radiatively generated by small quantum loop effects when they are broken at lower energies.  This is the approach we take to address the puzzling smallness of $\bar{\theta}$, or the strong CP problem as its called.  

The other approach is the well known anomalous $U(1)_{PQ }$ symmetry~\cite{1977PhRvL..38.1440P} whose spontaneous breaking gives rise to  the axion~\cite{PhysRevLett.40.223,PhysRevLett.40.279}. Quoting from Reference~\cite{Witten:2017hdv}: 	 ``The axion is a missing link to confirm the idea that 
 ``symmetries are only there to the extent that they are required by gauge symmetry.""

Since axions have not so far been discovered, and the idea is not confirmed, we take seriously the possibility that the strong CP problem points towards P and CP symmetric laws of nature~\cite{Nelson:1983zb, PhysRevLett.53.329, PhysRevD.41.1286, Kuchimanchi:2010xs}.

\textsl{\textbf{Parity (P) -- }} The Pati-Salam model~\cite{PhysRevD.10.275, PhysRevD.11.566} based on $SU(4)_C \times SU(2)_L \times SU(2)_R$ not only unified quarks and leptons into the same $SU(4)_C$ multiplet, but also it showed the way to achieve parity between the left and right handed weak interactions, without duplication of fermion families by mirror generations, and with the usual fermions acquiring masses from an $SU(2)_L \times SU(2)_R$ bi-doublet Higgs.  However the path to $SO(10)$ unification is through its sub-group $SU(4)_C \times SU(2)_L \times SU(2)_R \times C$ which involves charge conjugation rather than $SU(4)_C \times SU(2)_L \times SU(2)_R \times P$ that has parity.    

The popular  left-right symmetric model~\cite{PhysRevD.10.275, PhysRevD.11.566, Senjanovic:1975rk}  based on $SU(3)_C \times SU(2)_L \times SU(2)_R \times U(1)_{B-L} \times P$ undoes the quark-lepton unification aspect of the Pati-Salam group while promoting 
the idea of parity, 
and can be viewed as being in the different direction of restoring discrete spacetime symmetries, and therefore in this work we do not expect that the CP phases in the lepton and quark sectors will necessarily be similar. 

An early attempt was made to solve the strong CP problem in the left right symmetric model~\cite{Mohapatra:1978fy} by imposing more symmetries, and \textit{more scalars} to break them, but did not work as $ \bar{\theta} \sim \delta_{CKM}$ are either both or neither generated.  

The strong CP problem was later solved in the left-right symmetric model with the bi-doublet Higgs and P, by adding one vectorlike heavy quark family~\cite{Kuchimanchi:2010xs} (\textit{rather than more scalars}), whose dimension 3 mass terms with the usual light quarks break CP softly. 
 We will recall this solution in Section~\ref{sec:SCPLR} before adding an extra symmetry to bring it to the Nelson-Barr form that has so far only been associated with \textit{spontaneous} CP violation, introduced later in this section.  

\textsl{\textbf{Mirror Parity--}}  There are solutions using P as a mirror symmetry~\cite{PhysRevD.41.1286, D_Agnolo_2016, Craig:2020bnv}, which as  envisaged by Lee and Yang~\cite{PhysRev.104.254} requires duplication of the existing 3 fermion families by their mirror counterparts which form 3 heavy families of quarks and leptons.  
While the fermions are duplicated, the $SU(3)_C$ group may~\cite{bonnefoy2023colorful} or may not~\cite{PhysRevD.41.1286, D_Agnolo_2016, Craig:2020bnv} be. 

Surprisingly, we find in Section~\ref{sec:mpnb} that 
there is an \textit{automatic} Nelson-Barr solution on imposing CP,  in models with  $SU(2)_L \times SU(2)_R$ group containing respective Higgs doublets, where mirror parity is generally imposed. This opens a new approach with this group, where CP by itself solves the strong CP problem without needing any other symmetries, and also without requiring all 3 mirror generations. 

Mirror parity (which uses doublet Higgses rather than the bidoublet Higgs), like the SM,  does not necessarily anticipate that neutrinos have masses -- while $\nu$ masses can be radiatively generated from bare mass terms involving exotic charged leptons~\cite{Babu:2022ikf}, these terms can be set to zero using symmetries.

That neutrinos have masses  however is a prediction of the minimal left right symmetric model with bi-doublet Higgs, that is now borne out by experiments. The Dirac mass term for the $\nu_\tau$  generated on one-loop RGE running from the $\tau^-$, top and bottom Yukawa couplings is many orders of magnitude too large~\cite{Kuchimanchi:2017bfm} in this model, and therefore the neutrinos must have Majorana masses via the seesaw mechanism.  These predictions provide the motivation for our work with bi-doublet Higgs and parity.  

Note that in literature the same words parity and left right symmetry are often used even for mirror parity. Also the words generalized parity are sometimes used.

\textsl{\textbf{CP (or equivalently T) --}} We now turn to the solutions of the Strong CP problem using CP.  Arguably next only to axions in popularity, is the Nelson-Barr solution~\cite{Nelson:1983zb, FF1984165, PhysRevLett.53.329,  Bento:1991ez, Branco:2003rt}, that requires a heavy vectorlike quark(s)  which has (have) Yukawa couplings with the usual quarks via scalars whose vacuum expectation values (VEVs) break CP spontaneously. To prevent $\bar{\theta}$ from being generated at the tree level, an additional global symmetry is also imposed (which we find in Section~\ref{sec:mpnb} can be accidental or automatic). 

Obtained in the above way, NB mechanism splits the mass terms involving the heavy quark into dimension 3 terms that conserve CP and terms that are generated by Yukawa couplings and CP breaking VEVs, and requires these two types of terms to be at more or less the same scale.  This requirement of coincidence of unrelated scales is considered the most unsatisfying feature of NB models \cite{Dine:2015jga, Valenti_2021}.  

Moreover NB mechanism that involves $SU(2)_L$ doublet vectorlike heavy quarks obtained in the usual manner is almost ruled out~\cite{Vecchi2017} as the radiatively generated two loop strong CP phase is larger than the bounds from neutron edm experiments. Therefore most NB models only involve $SU(2)_L$ singlet up and down type heavy vectorlike quarks. 

\textsl{\textbf{In this work}} we arrive at the Nelson-Barr mechanism in a completely novel way.  We begin with the left right symmetric model containing the bidoublet Higgs, and use P to set $\bar{\theta} = 0$ (instead of CP that is used in NB models).  The strong CP problem is solved by breaking CP softly by dimension 3 mass terms following Reference~\cite{Kuchimanchi:2010xs}.   An additional symmetry brings the quark mass matrices to the Nelson-Barr form.  The CKM phase is generated through the Nelson-Barr mechanism. Since CP conserving and violating terms can be of the same type (whether CP is broken softly, or as discussed in Section~\ref{sec:spon} spontaneously), the problem of coincidence of scales does not in occur the NB solution we obtain in Section~\ref{sec:nbwithp}.

The critical non-decoupling type radiative corrections to $\bar{\theta}$ vanish in one loop, as is expected due to the NB form, and are generated in two loops. Further if P breaks at the scale $v_R$ below the soft CP breaking scale $M$,  it protects $\bar{\theta}$ between $v_R$ and $M$ and the two loop corrections are suppressed by the factor $(v_R/M)^2$, and also decouple.  Therefore a heavy vectorlike $SU(2)_L$ doublet, that is ruled out through the traditional CP based NB solution,  can be present in our P based NB mechanism,  if its mass $M > v_R$  due to the suppression of radiative corrections. The  two loop corrections to  $\bar{\theta} \sim 10^{-9} (v_R/M)^2$  are discussed in Section~\ref{sec:radnb}. 

Interestingly if $M < v_R$ while the NB mechanism generates an irreducible $\bar{\theta \sim 10^{-9}}$ that is now not suppressed, we do not need to impose the NB form, while still retaining the strong CP solution due to P.  In this case we can generate $\delta_{CKM}$ without the  NB mechanism, and surprisingly we find that a much lower irreducible contribution  $\bar{\theta} \sim 10^{-10~to~-13} ln (v_R/M)$ is generated in one-loop (and the two loop corrections are equally small).  This region of parameter space is entirely testable by current and future neutron EDM experiments. The irreducible contribution is discussed in Section~\ref{sec:radnonb}      
    
Leptonic CP violation is not  generated in the minimal model of this work (nor in~\cite{Kuchimanchi:2010xs}), providing a way to test the restoration of discrete spacetime symmetries, even if $v_R$ and $M$ are at very high scales from multi-Tev to $10^{15} GeV$, or more. The global fit to data from neutrino oscillation experiments is currently consistent with this prediction, discussed in Section~\ref{subsec:abs}. 

While the absence of leptonic $\delta_{CP}$ has been discussed before~\cite{Kuchimanchi:2012xb,Kuchimanchi:2012te} in the context of Reference~\cite{Kuchimanchi:2010xs}, and in ~\cite{Kuchimanchi_2015}, in this work we also discuss the challenges in a non-minimal model that can generate leptonic CP phases, in Section~\ref{subsec:pres}. 
 
The rest of the paper is organized as follows.  Section~\ref{sec:spon} shows how to introduce singlets without spoiling the strong CP solution,  so as to have spontaneous CP violation. Section~\ref{sec:mpnb} is a new approach to the strong CP problem with $SU(2)_L \times SU(2)_R$ group and respective Higgs doublets (rather than bi-doublet), where we find that there is an automatic Nelson-Barr solution without requiring mirror parity. Section~\ref{sec:minsm} is the minimal way to achieve the Nelson-Barr mechanism with softly broken CP symmetry and a real scalar singlet.   Section~\ref{sec:nr} is a brief discussion on non-renormalizable terms and in Section~\ref{sec:conc} we present the conclusions.    


\section{Strong CP solution and Nelson-Barr mechanism with P}
\label{sec:nbwithp}
We first recall in Subsection~\ref{sec:SCPLR}, the solution to the strong CP problem in the Left-Right symmetric model with Parity~\cite{Kuchimanchi:2010xs}.   We then impose in Subsection~\ref{sec:nbform} a softly broken  symmetry to bring the quark mass matrices to the Nelson-Barr form.  

\subsection{Strong CP solving left right symmetric model with bi-doublet Higgs} 
\label{sec:SCPLR}
We begin with the Left-Right symmetric model~\cite{PhysRevD.10.275,*PhysRevD.11.566,Senjanovic:1975rk} based on $SU(3)_c \times SU(2)_L \times SU(2)_R \times U(1)_{B-L} \times P$ with the Higgs sector consisting of scalar triplets $\Delta_R$ and its parity partner $\Delta_L$ and a bi-doublet $\phi$ 
(for the Higgs potential please see for example~\cite{Duka:1999uc,ZHANG2008247}).    

Following Reference~\cite{Kuchimanchi:2010xs}, in addition to the usual three quark families $Q_{iL}$ and $Q_{iR}$, which are doublets of $SU(2)_L$ and $SU(2)_R$ respectively,  we add a fourth vector like quark doublet family, whose left and right handed components $Q_{4L}$ and $Q'_{R}$ are both $SU(2)_L$ doublets.    Due to parity there is also the corresponding $SU(2)_R$ doublet vector-like family with components:  $Q_{4R}$, $Q'_{L}$. In the minimal version there are the usual 3 generations of leptons $L_{iL}, L_{iR}$ and there is no vector-like lepton family.

Note that the usual right handed $SU(2)_L$ singlet quarks and leptons, are all in the corresponding $SU(2)_R$ doublets $Q_{iR}$ and $L_{iR}$, and we automatically have the right-handed neutrinos in $L_{iR}$. 

The scalar and fermion content is displayed in table~\ref{tab:matterta} and here after we assume $i$ runs from $1$ to $4$ for the quarks to include the usual 3 light chiral and the $4^{th}$ normal chiral component of the vector like quark family. While for the leptons, $i$ runs from $1$ to $3$. 

\begin{table}[h]
\centering
\begin{tabular}{cc}
\hline 
& $SU(3)_C \times SU(2)_L \times SU(2)_R \times U(1)_{B-L}$ \\  \hline \\ 
$\Delta_L$ & $(1,3,1,2)$ \\ 
$\Delta_R$ & $(1,1,3,2)$ \\ 
$\phi$ & $(1,2,2,0)$ \\ \\ \hline \\
$Q_{iL}, Q'_R $ & $(3,2,1,1/3)$ \\ 
$Q_{iR}, Q'_L $ & $(3,1,2,1/3)$ \\ 
$L_{iL} $ & $(1,2,1,-1)$ \\ 
$L_{iR} $ & $(1,1,2,-1)$ \\ \\ \hline

\end{tabular}
\caption{Scalar and fermion content of the minimal strong CP solving LR model.  Note that i = 1 to 4 for quarks and i = 1 to 3 for leptons}
\label{tab:matterta}
\end{table}

Parity  reflects $(x,t) \rightarrow (-x,t)$, sets $\theta_{QCD} = 0$, exchanges the $SU(2)_L$ and $SU(2)_R$ gauge bosons, and  
\begin{equation}
Q_{iL} \leftrightarrow Q_{iR}, Q'_{L} \leftrightarrow Q'_{R}, L_{iL} \leftrightarrow L_{iR}, \Delta_L \leftrightarrow \Delta_R, \phi \rightarrow \phi^\dagger.
\label{eq:eta}
\end{equation}
$SU(2)_R \times U(1)_{B-L} \times P \rightarrow U(1)_Y$,  when the neutral component of $\Delta_R$ picks up a VEV $\sim v_R > v_{wk}$. We do not make any assumptions on the scale of  $v_R$ and it can be from multi-TeV scale~\cite{PhysRevD.101.035036,Kuchimanchi:2017bfm} to the canonical value of seesaw scale $\sim 10^{13~to~14} GeV$ set by neutrino's Dirac Yukawa couplings being like that of the top quark, or even above.    

The SM group is broken by the VEVs $\left<\phi^0_1\right> \equiv \kappa_1$ and $\left<\phi^0_2\right> \equiv \kappa_2$  of the bidoublet $\phi$ (with the weak scale $v_{wk}^2 = |\kappa_1|^2 +|\kappa_2|^2$).  $\phi$ has two SM Higgs doublets labeled below by subscripts 1 and 2,  and can be represented by the matrix 
\begin{equation}
\phi = \left(
\begin{array}{cc}
\phi^0_1 & \phi_2^+ \\
\phi^-_1  & \phi^0_2
\end{array}
\right)
\label{eq:phi}
\end{equation}

When $\phi$ picks up VEVs $\kappa_{1,2}$, the Yukawa couplings in equation~(\ref{eq:yukawata}) with $\tilde{\phi} = \tau_2 \phi^\star \tau_2$, generate the up and down sector quark mass matrices $M_u$ and $M_d$. Since under P, $\phi \rightarrow {\phi}^\dagger$, a complex phase for $\left<\phi^0_2\right> \equiv \kappa_2$ ($\kappa_1$ can always be chosen to be real) breaks $P$  as well as $CP$, and will generate $\bar{\theta} = Arg Det (M_u M_d)$ at the tree level. Therefore we impose CP on all dimensionless quartic couplings of the Higgs potential and Yukawa couplings. 

CP ensures that the sole parameter $\alpha_2$ of the Higgs potential that could have been complex, stays real (note that this P symmetric term $\alpha_2 Tr (\tilde{\phi}^\dagger \phi \Delta^\dagger_R \Delta_R  + R, \phi \rightarrow L, \phi^\dagger) + h.c.$ is automatically absent if there is supersymmetry~\cite{Kuchimanchi:1995rp, Mohapatra:1995xd},  but in this work we are considering non-supersymmetric models and therefore impose CP).   There is a soft mass term in the Higgs potential between the two SM doublets of the bidoublet, $\mu^2 \tilde{\phi}^\dagger \phi + h.c.$ and we note that P ensures that $\mu^2$ is real, without needing CP. This parameter stays real due to P even after we include dimension 3 mass terms in equation~(\ref{eq:masstermsta}) that break CP softly.  There are therefore no complex parameters in the scalar potential and consequently the bidoublet VEVs, obtained by minimizing the Higgs potential, are real (conserve P and CP) at the tree level although P is broken by the VEV of $\Delta_R$.

We note that the Yukawa terms can be written as,
 {
\begin{equation}
\underline{h}_{ij} \ \bar{Q}_{iL} \ \phi \ Q_{jR} \ + \  \tilde{\underline{h}}_{ij} \ \bar{Q}_{iL} \ \tilde{\phi} \ Q_{jR} \   + \ h.c,
\label{eq:yukawata}  
\end{equation}}
 
where $\underline{h}_{ij}, \tilde{\underline{h}}_{ij}$ are Hermitian due to P and real due to CP and $i,j$ run from 1 to 4.  Note that one of the SM doublets in the bidoublet $\phi$ picks up a mass $\sim v_R$ and decouples from low energy physics. And there is only one SM Higgs doublet below $v_R$ whose Yukawa couplings
\begin{eqnarray}
 \underline{h}^u_{ij} v_{wk} & = & \underline{h}_{ij} \kappa_1 + \tilde{\underline{h}}_{ij} \kappa_2, \nonumber \\
\underline{h}^d_{ij} v_{wk} & = & \underline{h}_{ij} \kappa_2 + \tilde{\underline{h}}_{ij} \kappa_1 
\label{eq:yukta}
\end{eqnarray}
for the up and down sector are obtained from the above real and Hermitian Yukawas and real Higgs VEVs $\kappa_1, \kappa_2$.    

We use the convention of Reference~\cite{Kuchimanchi:2010xs} of underlining the real Yukawa couplings in the symmetry basis above the CP breaking scale.

Soft CP breaking is introduced by the direct dimension 3 mass terms between the light and heavy quarks and generates CP violation, while respecting Parity.

\begin{equation}
M_i \ \bar{Q}_{iL}  {Q'}_{R} \ + \ M_i^{\star} \  \bar{{Q'}}_{L}  Q_{iR} \ + \ h.c.
\label{eq:masstermsta}
\end{equation}
where the sum over repeated index i goes from 1 to 4 and we have imposed P.

Note that the up and down quark mass matrices $M_u, M_d$ obtained from equations~(\ref{eq:yukta}) and (\ref{eq:masstermsta})  while CP violating, are Hermitian (respect P) and therefore $\bar{\theta}= Arg  Det(M_u M_d) =0$ at the tree level, thereby solving the strong CP problem. Below the mass of the heavy vector like quarks, we integrate them out by going to the heavy quark mass basis, and the CP violation shows up as the CKM phase $(\delta_{CKM})$ in the SM Yukawa couplings $h^u$ and $h^d$, which are Hermitian at the tree level. More details of this solution can be found in Reference~\cite{Kuchimanchi:2010xs}.    

Note that the the Yukawa couplings in equation~(\ref{eq:yukawata}) do not have the Nelson-Barr form.  Moreover there are no scalar singlets whose VEVS contribute to $M_i$ of equation~(\ref{eq:masstermsta}). All the mass parameters $M_i$ (with $i=1$ to $4$) are complex and are on an equal footing, since they all break softly the accidental discrete symmetry ($Q'_L, Q'_R \rightarrow -Q'_L, -Q'_R$) of the rest of the terms of the Lagrangian. Note that this symmetry argument applies even with the Nelson-Barr form obtained in the next subsection.    Therefore the problem of needing the coincidence of unrelated scales between CP conserving dimension 3 mass terms and CP violating terms generated by scalar VEVs, that is generally present~\cite{Dine:2015jga, Valenti_2021} in Nelson-Barr solutions, is not there in our approach.

We now impose an additional softly broken symmetry to obtain the Nelson-Barr form of the mass matrices.

\subsection{Obtaining Nelson-Barr form}
\label{sec:nbform}
We impose a chiral symmetry under which $Q_{4L} \rightarrow e^{i\alpha} Q_{4L}$. This sets $\underline{h}_{4j}, \underline{\tilde{h}}_{4j} =0$ in equation~(\ref{eq:yukawata}),  and  
\begin{equation}
\underline{h}^{u,d}_{4j} = \underline{h}^{u,d}_{j4} =0
\label{eq:yukzero}
\end{equation}
 for all $j$ in equation~(\ref{eq:yukta}), since the Yukawa matrices are Hermitian.  $M_4$ in equation~(\ref{eq:masstermsta}) breaks this symmetry softly.

We can now make equal left and right chiral rotations $Q_{4L,4R} \rightarrow e^{i\beta}Q_{4L,4R}$ and choose $\beta$ so that $M_4$ is made real.  

Using the up and down components of the doublets $Q^T_{iR} \equiv (u_{iR} \ d_{iR})^T$ and $Q'^T_{R} \equiv (u'_{R} \ d'_{R})^T$ (and likewise for $R \rightarrow L$), the Hermitian up (and similarly down) sector quark mass terms can be written as
\begin{math}
\left({{\bar{u}}_{L}, \bar{u}'_{L}}\right) \ M_u \ 
\left(\begin{array}{c}
{u}_{R}  \\ 
u'_{R}  \ 
\end{array}\right)
\end{math}, 
where $u_R$ is shorthand for $u_{iR}$, and is a $4 \times 1$ column vector in family space and contains the usual 3 light chiral generations as well as the corresponding 4th chiral component of a heavy vector like generation (and likewise $u_L$).  Up and down quarks mass matrices $M_u$ and $M_d$ obtained from equations~(\ref{eq:yukta}), (\ref{eq:masstermsta}) and~\ref{eq:yukzero} now have the Nelson-Barr form and can be written as:

\begin{equation}
M_{u,d} =
\left(\begin{array}{ccc|c|l}      
\ \  & \ \  & \ \   & 0 &  M_1 \\       
\ \  & \underline{h}^{u,d} v_{wk}  & \ \ & 0 & M_2 \\
\ \ & \ \   & \ \  &  0  & M_3 \\
\hline
0  & 0  & 0  &  0  & M_4 \\
\hline
M_1^\star & M_2^\star & M_3^\star & M_4^\star &  0
\end{array}\right)
\label{eq:massmatrixta}
\end{equation}

with $M_4$ real and $\underline{h}^{u,d}$ now being the respective $3 \times 3$ Yukawa matrices of equation~(\ref{eq:yukta}) with elements $\underline{h}^u_{ij}$ and $\underline{h}^d_{ij}$ with $i,j=$1 to 3.  Without loss of generality, to ease the calculations, the real Yukawa couplings of terms  $\bar{Q}'_L \phi^\dagger Q'_R$ and $\bar{Q}'_L \tilde{\phi}^\dagger Q'_R$ that contribute to the 5th row's 5th column have been set to zero by imposing a chiral symmetry $Q'_R \rightarrow e^{i\alpha} Q'_R$ broken softly by dimension 3 mass terms $M_i$.

Note that $M_u$ matrix is $5 \times 5$ is because there are two heavy vectorlike up quarks -- one which is a heavy $SU(2)_L$ singlet (from the $SU(2)_R$ doublet), and the other is from the heavy $SU(2)_L$ vectorlike doublet.  If we remove the $4th$ row and $5th$ column from the above matrix, we will recognize what remains as being the familiar Nelson-Barr form with just the heavy singlet up quark. On the other hand if we remove the $4th$ column and $5th$ row what remains is the form with the heavy doublet up quark.  Likewise for the down sector.


There are two reasons now why the determinants of $M_{u}$ and $M_d$ are real at the tree level. One reason is because these matrices are Hermitian due to P. The other because they are of the Nelson-Barr form and complex parameters are multiplied by zeros while evaluating the determinant.       
  
We have obtained the Nelson-Barr form by using P rather than CP  to set $\bar{\theta}$ to zero at the tree-level.  The net result is that there is no need to introduce iso-singlet scalars of the NB type that have Yukawa couplings with the heavy quarks, and  the model is free from the coincidence of scale problem~\cite{Dine:2015jga, Valenti_2021} between CP violating terms (from scalar singlet vevs that break CP) and CP conserving mass terms (that don't couple to the singlets).

Below the mass 
\begin{equation}
M^2 = \sum\limits_{i=1}^4 |M_i|^2
\end{equation} 
of the heavy quarks (which set the soft-CP breaking scale $M$),  we can go to their mass basis by a unitary transformation so that the mass matrices $M_{u}$ and $M_{d}$ become~\cite{Kuchimanchi:2010xs}:
\begin{equation}
M_{u,d} =
\left(\begin{array}{cccc|l}      
\ \  & \ \  & \ \   &  &  0 \\       
\ \  & h^{u,d} v_{wk}  & \ \ &  & 0 \\
\ \ & \ \   & \ \  &   & 0 \\
 &   &   &   & M \\
\hline
0 & 0 & 0 & M &  0
\end{array}\right)
\label{eq:diagmassmatrixta}
\end{equation}
where $h^{u}$ and $h^{d}$ are complex $4 \times 4$ Hermitian Yukawa matrices, whose light (upper left) $3\times 3$ sector are the SM Yukawa couplings, and $M$ is real.

$\delta_{CKM}$ is thus generated through the Nelson-Barr mechanism.  Note the important difference from the way it was generated in the example given in the Appendix of Reference~\cite{Kuchimanchi:2010xs}, where the Yukawa couplings (or quark mass matrices) were not of the NB form. Specifically note the critical role played by the Yukawa coupling $\underline{h}_{4c}$ (that is, $\underline{h}^u_{42}$) to generate $\delta_{CKM}$  in that example, which is now zero in the NB form of equation~(\ref{eq:massmatrixta}).  Therefore the radiative corrections to $\bar{\theta}$ worked out in that example are also different, and we now  discuss the loop corrections within the NB framework.

\section{Radiative corrections to $\bar{\theta}$}
\label{sec:rad}

\subsection{Corrections with the Nelson-Barr mechanism}
\label{sec:radnb}
Radiative corrections in Nelson-Barr models have comprehensively been estimated in References~\cite{Vecchi2017}, building on the work of Nelson~\cite{FF1984165}.  

There are generally two types of corrections, ones that are suppressed by the mass scale of the heavy quarks are of the decoupling type.  However there are also the more dangerous non-decoupling corrections that are not suppressed. 


We will first consider the case with $M < v_R$ with the heavy quarks  just below the parity breaking scale.  Below the scale $M$, we would like to integrate out the heavy quarks and consider just the effective SM theory. 

As noted by Nelson~\cite{FF1984165} in the mass basis of the heavy quarks, the Yukawa couplings between the heavy and  light quarks can generate radiative corrections to the light quark mass matrices on electro-weak symmetry breaking. These corrections can come from two loop diagrams involving the SM Higgs and at least one of the heavy quarks in the internal lines. These 2 loop diagrams are similar to the diagrams that would renormalize the Yukawa couplings (and involve the term $h^uh^{u\dagger}h^dh^{d\dagger}$), while the one-loop contributions to $\bar{\theta}$ vanish due to the Nelson-Barr form.  

Naively, we may think that the corrections to $\bar{\theta}$ from the heavy quarks would be suppressed by $\left<H^\dagger H\right>/M^2 \sim (v_{wk}/M)^2$, where $H$ is the SM Higgs doublet field.  However since the loop diagrams that renormalize the Yukawa couplings, are logarithmically divergent in the ultraviolet, when we pull out the factor $1/M^2$ from the internal fermion line that has the heavy quark (since we are interested in loop momenta $p^2$ less than $M^2$ while evaluating the Feynman integral), the remaining part of the integrand (now with one $p^2$ factor less in the denominator) is quadratically divergent.  When we integrate up to a mass scale say $M' \leq M$ this provides a contribution proportional to $(M'/M)^2$, so that  

\begin{equation}
\bar{\theta} \sim \left(1/16 \pi^2\right)^2 (M'/M)^2 Im Tr [ h^uh^{u\dagger}h^dh^{d\dagger}]_{3\times 3}
\label{eq:2loop}
\end{equation}
 where we have used square brackets and the subscript to convey that the trace is over the light (upper left) $3 \times 3$ sub-matrix, while $h^u$ and $h^d$ are the $4 \times 4$ Hermitian, complex Yukawa matrices in the mass basis (equation~(\ref{eq:diagmassmatrixta})) of the heavy quarks. Note that the reason the Trace is over the light $3 \times 3$ sub-matrix is because the Yukawa couplings involving the 4th (heavy) quark, $h^{u,d}_{i4}$ and $h^{u,d}_{4i}$, multiply $0$ when the determinant of $M_u$ (and $M_d$) is evaluated in the heavy quark mass basis of equation~\ref{eq:diagmassmatrixta}, and therefore radiative corrections to $h^{u,d}_{i4}$ do not contribute to $\bar{\theta}$ in the leading order. 

\emph{If $M < v_R$} then we have to integrate till the heavy quark mass scale and so $M'=M$ in equation~(\ref{eq:2loop}).  We substitute the values of $h^u$ and $h^d$ we obtain from $\underline{h}^{u,d}$ in equation~(\ref{eq:massmatrixta}) by using typical values of $|M_i/M_4| \sim 0.5$ (for i=2,3) and making a unitary transformation into the heavy quark mass basis of equation~(\ref{eq:diagmassmatrixta}).   We find on substitution of typical values that to obtain $\delta_{CKM} \sim 1$ with the known light quarks' masses and mixing angles:   
  \begin{equation}
\bar{\theta}_{M < {v_R}} \sim \left(1/16 \pi^2\right)^2 Im Tr [h^uh^{u\dagger}h^dh^{d\dagger}]_{3 \times 3} \sim 3 \times 10^{-10}
\label{eq:NB}
\end{equation}     

This is in agreement with $\bar{\theta} \sim (1/4\pi^2)^2 \lambda_C^2 Y_t^{2}Y_b^{2} \sim 5 \times 10^{-9}$ (where $Y_{t,b}$ are the top and bottom quarks' Yukawa couplings and $\lambda_C$ is the Cabibo angle), estimated in Reference~\cite{Vecchi2017} for NB mechanism via a vectorlike $SU(2)_L$ doublet quark, that our model contains.

As noted in Reference~\cite{Vecchi2017} the contributions to $\bar{\theta}$ from the up and down mass matrices don't cancel each other.  Since the placement of the internal heavy quark lines is in general different for different combinations of Yukawa couplings in the up and down sectors, we also do not expect cancellation between their contributions. 

\emph{On the other hand if $v_R < M$,} then parity protects $\bar{\theta}$ from  radiative corrections above $v_R$.  The second Higgs doublet in the bidoublet provides canceling contributions above its mass $\sim \alpha_3 v_R^2$.  Therefore $M'^2 = \alpha_3 v_R^2$ in equation~(\ref{eq:2loop}) and 
\begin{eqnarray}
\bar{\theta}_{{v_R} < M} & \sim & \left(1/16 \pi^2\right)^2 Im Tr [h^uh^{u\dagger}h^dh^{d\dagger}]_{3 \times 3} \ \alpha_3 \ (v_R/M)^2  \nonumber \\
& \sim & 10^{-9} \alpha_3 (v_R/M)^2
\label{eq:P}
\end{eqnarray}

where, as before, the trace is over the upper left $3 \times 3$ sub-matrix part of the product of the $4 \times 4$ Yukawa matrices. Note that $\alpha_3 \sim 0.1$ to $1$ is the real quartic coupling of the Higgs potential term $Tr(\phi^\dagger\phi \Delta_R^\dagger \Delta_R + R\rightarrow L, \phi \leftrightarrow \phi^\dagger)$.

The factor $(v_R/M)^2$ is consistent with treating $M$ as a cut off scale for new physics in the ultraviolet of the minimal LR model with P (without the heavy quarks), and including the effect of  the scale of the heavy quarks  by non-renormalizable terms. 
$\bar{\theta}$ can be generated when $\Delta_R$ acquires the VEV $v_R$ by non-Hermitian parameters of non-renormalizable terms such as $\bar{Q}_{iL} \phi Q_{jR} (\Delta_R^\dagger \Delta_R/M^2)$ $+ L \leftrightarrow R, \phi \rightarrow \phi^\dagger$, which decouple as $(v_R/M)^2$.  In our case, the pre-factor to $(v_R/M)^2$ that depends on the Yukawa couplings, as we calculated is $\sim 10^{-9}\alpha_3$. 

Note that such decoupling does not happen with just the SM because there is no symmetry in the SM that sets $\theta_{QCD}$  to zero. SM parameters (including $\bar{\theta}$) can get contributions that are not suppressed by the scale of new physics.   Therefore in just the SM without P, the radiative corrections of the NB mechanism generate an unsuppressed $\bar{\theta}$, which is also the case in equation~(\ref{eq:NB}), for $M < v_R$. 

Since $\bar{\theta}$ in equation~(\ref{eq:P}) can be acceptably small if $M$ is just a factor of $10$ or $100$ larger than $v_R$, the CP breaking scale does not have to be much larger than the $SU(2)_R \times U(1)_{B-L} \rightarrow U(1)_Y$ breaking scale. 

We have thus found an interesting resolution of the strong CP problem which provides an ultraviolet completion of the  minimal left-right symmetric model, with the Nelson-Barr mechanism.



\subsection{Corrections without Nelson-Barr}
\label{sec:radnonb}
The radiative corrections from the NB mechanism are too large if the heavy quarks are below the P breaking scale (that is, if $M < v_R$). $\bar{\theta} \sim 10^{-9}$ that is generated in equation~(\ref{eq:NB}) is \textsl{irreducible}~\cite{Vecchi2017} since the very couplings that are responsible for generating $\delta_{CKM}$ also generate the strong CP phase.  

Therefore we now remove the chiral symmetry that led to the NB form and consider the more general case~\cite{Kuchimanchi:2010xs} of the mass matrix for the case $M < v_R$.  Now the $\delta_{CKM}$ can be generated without the NB mechanism, from the phases in $M_i$ and using Yukawa couplings $\underline{h}^{u,d}_{4j} = \underline{h}^{u,d}_{j4}$ between the heavy and usual quarks as they no longer vanish. 

What we find is that even in this case there is an irreducible contribution, but surprisingly it is smaller and has an experimentally interesting value $ \bar{\theta}_{M<v_R} \sim  (10^{-10}~to~10^{-13}) ln (v_R/M)$. Corrections of this order were found in Reference~\cite{Kuchimanchi:2010xs} in one-loop, but they were not shown to be irreducible - in the sense that depending on the choice of the parameters they could also vanish. 

We will now find an irreducible correction $\bar{\theta}~\sim 10^{-13}ln(v_R/M)$ that doesn't vanish in one and two loops (and more likely contributions that are 10-1000 times higher), and therefore the entire parameter space of $M < v_R$ can be tested in the  future (earlier than later) by neutron EDM experiments. 


To obtain the above we work in a basis where the upper left $4 \times 4$ sub-matrix of the Hermitian matrix $M_u$ given by equations~(\ref{eq:yukta}) and (\ref{eq:masstermsta}) 
is diagonal, and we set some of the parameters to zero to obtain the smallest $\bar{\theta}$, for which we choose $M_u$ and $M_d$ of the form

  \[
		{M_{u} =
\left(\begin{array}{cccc|l}      
 0 &   &    & &  M_1 \\       
 & \underline{h}^{u}_{22} v_{wk}  & \ \ &  & 0 \\
 &   & \underline{h}^u_{33} v_{wk}  &   & 0 \\
 &   &   & \underline{h}^u_{44} v_{wk} & M_4 \\
\hline
M^\star_1 & 0 & 0 & M_4 & 0  
\end{array}\right)} 
\]

\begin{equation}
{M_{d} =
\left(\begin{array}{cccc|l}      
\underline{h}^d_{11} v_{wk} &  0 & \underline{h}^d_{13} v_{wk} & 0 &  M_1 \\       
0 & \underline{h}^{d}_{22} v_{wk}  & \underline{h}^d_{23} v_{wk} &  \underline{h}^d_{24} v_{wk}  &0\\
\underline{h}^d_{13} v_{wk} & \underline{h}^d_{23} v_{wk}   & \underline{h}^d_{33} v_{wk}  & 0  & 0 \\
0 &  \underline{h}^d_{24} v_{wk} & 0  &  0 & M_4 \\
\hline
M^\star_1 & 0 & 0 & M_4 &  0
\end{array}\right)}
\label{eq:smallmat}
\end{equation}

Note that since the above matrices are Hermitian and only $M_1$ is complex, there are just as many non-zero Yukawa couplings as the light quarks masses and mixing angles.  The phase in $M_1/M_4$  generates $\delta_{CKM}$ (and the ratio with $\underline{h}^d_{24}$ the Cabibo angle) when we go to the mass basis of the heavy quarks with a unitary transformation in the $1-4$ plane.  Non-zero contributions we find for $\bar{\theta}$ with the above minimal choice of non-zero parameters is irreducible.  

We choose $|M_1/M_4| \sim 0.7$ (much smaller values tend to give a higher $\bar{\theta}$ as $\underline{h}^d_{24}$ would have to be higher to get the Cabibo angle).  All the remaining non-zero terms in the above two matrices are determined from the light quark masses, mixing angles, and $\delta_{CKM}$. We find below an irreducible $\bar{\theta} \sim 10^{-13}$ with this choice of parameters.

But before that, to understand the plausible physics that could generate roughly the above form of the matrices, we note that $M_2 = M_3 =0$ can be obtained by imposing  the symmetry under $Q_{1L,4L} \rightarrow -Q_{1L,4L}, \ Q'_R \rightarrow -Q'_R$  on Yukawa couplings and mass terms in~(\ref{eq:yukawata}), (\ref{eq:yukta}) and (\ref{eq:masstermsta}). The Yukawa couplings involving the first and fourth generation break this symmetry approximately due to the smallness of $\underline{h}^{u,d}_{1i}, \underline{h}^{u,d}_{i4} \sim 10^{-5 \pm 1}$ in $M_{u,d}$ of equation~(\ref{eq:smallmat}). Therefore $|M_2|, |M_3|$ that also break this symmetry approximately, can be smaller than $M_4$ by a similarly small factor, which justifies our setting them to zero in the leading order.   

We now go into the mass basis of the heavy quarks by bringing the mass matrices to the form in equation~(\ref{eq:diagmassmatrixta}) and obtain the complex Hermitian Yukawa matrices $h^u$ and $h^d$ (now written without the underline).

Since $M < v_R$, below the $P$ breaking scale, $h^{u,d}$ obtain non-Hermitian corrections $\Delta h^{u,d}$ on RGE running and as found in Reference~\cite{Kuchimanchi:2010xs}  generate:
\begin{equation}
\bar{\theta}  \sim  Im Tr ([h^d]^{-1}_{3 \times 3} [\Delta h^d]_{3 \times 3}) 
\label{eq:thetadown}
\end{equation}
with
\begin{equation}
 [\Delta h^d]_{3 \times 3} \sim   [ h^u h^{u\dagger}h^d ]_{3\times 3} ln \left(v_R/M\right) / \left(16 \pi^2\right)
\label{eq:Delta}
\end{equation}
where $h^u$ and $h^d$ inside the square brackets are the $4 \times 4$ complex Hermitian Yukawa matrices, and having evaluated the resultant matrix within any square brackets,  we drop its $4th$ row and $4th$ column to obtain the corresponding $3 \times 3$ matrices outside the square brackets.  The inverse is taken after obtaining the $3 \times 3$ submatrix, and therefore has been indicated outside the square brackets.

Note that in the above equation, $h^u h^{u\dagger}$ is from the beta function of the one loop RGE for $h^d$.  Therefore $h^uh^{u\dagger}h^d$ is the Yukawa factor contributing to $\Delta h^d$. 

There is also a contribution from RGE running of $h^u$ that is obtained with $u \leftrightarrow d$ in equations~(\ref{eq:thetadown}) and (\ref{eq:Delta}).

Using the form of $h^d$ and $h^u$ in equation~(\ref{eq:smallmat}), with their values  determined  by masses and mixing of the usual light quarks (with $|M_1/M_4| \sim 0.7$), we find from equations~(\ref{eq:thetadown}) and (\ref{eq:Delta}) an irreducible 
\begin{equation}
\bar{\theta}_{{M < v_R}_{irred.}} \sim 10^{-13} ln(v_R/M)
\label{eq:thetaone}
\end{equation}. 
While the contribution from $\Delta h^u$ vanishes.  

Note that using equations~(\ref{eq:smallmat}), (\ref{eq:thetadown}) and (\ref{eq:Delta}) the above contribution can be understood as $ (1/16\pi^2)Y^2_u \lambda^2_C \sim 10^{-13}$ where $Y_u \sim 2 \times 10^{-5}$ is the Yukawa coupling of the SM up quark and $\lambda_C$ is the Cabibo mixing angle. 

This irreducible contribution was missed in the example in the Appendix of Reference~\cite{Kuchimanchi:2010xs} as only a single one-loop contribution (which could have vanished) was evaluated, by providing non-zero values for some of the Yukawa couplings that could have been set to zero. Note that in that example a $\underline{h}^d$ diagonal basis was used, and so the corresponding contribution worked out was from $\Delta h^d$ that could have vanished. 

We also find that contributions from the two loop beta function term $h^{u} h^{u\dagger} h^d h^{d\dagger}$, in place of  $h^{u} h^{u\dagger}$ in equation~(\ref{eq:Delta}), with an additional loop factor of $1/16\pi^2$, are of a similar order of magnitude as~(\ref{eq:thetaone}), while all other two loop corrections  are several orders of magnitude smaller.  

Interestingly, the one and two loop corrections under RGE running evaluated using equations~(\ref{eq:thetadown}) and (\ref{eq:Delta}) vanish if  $h^u$ and $h^d$ \textit{are obtained from the Nelson-Barr form} (equations~(\ref{eq:yukzero}) and (\ref{eq:massmatrixta})). This makes sense as it is easy to see from the Yukawa and mass terms that the Nelson-Barr form is invariant under RGE running at scales above the mass $M$ of the heavy quarks. Therefore $\bar{\theta}$ is not generated under RGE running between $v_R$ and $M$.   The radiative corrections in the NB case arise at the scale  $M$, since terms that would cancel for RGE running above the scale $M$, no longer cancel at and below the scale $M$, as the heavy mass  cannot now be taken to be zero.  

The two loop term in equation~(\ref{eq:NB}) is one such term (occurring in $ Im Tr([ h^u h^{u\dagger}  h^d h^{d\dagger}  h^d]_{3\times 3}[h^d]^{-1}_{3 \times 3})$ when the upper left $3\times 3$ part of $h^d$ and $[h^d]^{-1}_{3 \times 3}$ multiply each other) that doesn't get canceled for the Nelson-Barr form at the scale $M$ (but is canceled by the remaining terms in the Trace, above the scale $M$, for RGE running).  

We therefore need to also check that $h^u h^{u\dagger} h^d h^{d\dagger} $ taken by itself, generates from equation~(\ref{eq:NB}) only a negligibly small $\bar{\theta}$ for the Yukawa matrices obtained from the form in equation~(\ref{eq:smallmat}), that  we considered in this sub-section. And this turns out to be the case.

One loop contributions that are 10-1000 times larger than in equation~\ref{eq:thetaone} are generated (interestingly those from $\Delta h^u$ are more dominant), when we turn on the Yukawa couplings and $M_i$ that we set to zero to obtain the form in equation~(\ref{eq:smallmat}). Moreover the logarithmic factor could be order 1-10.  Therefore we expect $\bar{\theta} \sim 10^{-13}~to~10^{-10} ln (v_R/M)$ is generated and will be found sooner than later by the neutron EDM experiments.

Since the problem of coincidence of scales is not there in our model, and since there are no scalars that are usually associated with NB mechanism, the heavy quarks can  be naturally light and well separated from the Planck scale, and potentially also be at the current collider scales. The form in equation~(\ref{eq:smallmat}) suggests that the heavy quarks couple very weakly to the usual quarks.  Therefore if a neutron EDM corresponding to  $\bar{\theta} \geq 10^{-12}$ is found, it may be worthwhile to look for such weakly coupled heavy quarks in the current and future colliders.
 
For the case where $v_R < M$, P protects $\bar{\theta}$ and the radiative contributions are suppressed by the factor $(v_R/M)^2$ as in the previous sub-section. The pre-factor will depend on the couplings in equation~\ref{eq:smallmat}, where we no longer need to restrict some Yukawa couplings and $M_i$ to be zero or very small.  

\section{Leptonic CP violation \newline or absence thereof }
\label{sec:lep}

\subsection{Absence in minimal model}
\label{subsec:abs}
In the Nelson-Barr mechanism with the SM gauge group and 3 right handed neutrinos, the NB scalar singlets whose VEVs generate $\delta_{CKM}$ in the quark sector, would also have Majorana type Yukawa couplings with $\nu_R$ (which is a SM singlet)  and generate the CP phase in the leptonic sector. In that way, it has been argued that there is a common origin of CP phases in both sectors~\cite{Branco:2003rt}.  

However in the left-right model, since the right handed neutrinos are $SU(2)_R$ doublets, they get their Majorana masses through Yukawa couplings with $\Delta_R$ whose VEVS do not violate CP. Also we imposed CP to ensure that the VEVs of the bidoublet $\phi$ are real, which was needed to solve the strong CP problem.  Therefore the VEV of $\Delta_L$ which is generated from real Higgs potential terms and real VEVs of $\Delta_R$ and $\phi$ is also real. The left or right handed neutrinos cannot pick up any CP violating Dirac or Majorana masses with the minimal particle content of Table~\ref{tab:matterta}.       


This is great for testing the model since, all leptonic CP phases including the Dirac phase $\delta_{CP}$ of the PMNS matrix must be zero or $\pi$. Global fits to data from current neutrino experiments are consistent with $\delta_{CP} = \pi$ to within one sigma~\cite{deSalas:2020pgw, Esteban:2020cvm} for normal ordering of neutrino masses.  $Sin \delta_{CP} = 0$  is a prediction that the next-generation neutrino experiments such as DUNE and Hyper-Kamiokande will be testing with much greater sensitivity. 


Before we proceed to the non-minimal model, it was noted in~\cite{Kuchimanchi:2012te} while discussing the above absence, that even in the case of Nelson-Barr solution in the SM and 3 right handed-neutrinos, we can also assign the symmetry \textit{required } to obtain the NB form such that the NB scalars do not couple to $\nu_R$, and then no leptonic CP violation is generated.  While in Reference~\cite{Branco:2003rt}, it was assumed that leptonic CP is violated, and its possible absence was not considered.

\subsection{Challenges for non-minimal model}
\label{subsec:pres}
In order to generate CP phases in the leptonic sector, we add to the particle content of Table~\ref{tab:matterta} a heavy vector like $SU(2)_L$doublet lepton family $L_{4L}$ and $L'_R$ and its parity counterpart $L_{4R}$ and $L'_L$ which is an $SU(2)_R$ doublet family (and $SU(2)_L$ singlets). 

Unlike for the heavy quarks, note that  there are two physically distinct choices for the heavy leptons to transform under P~\cite{Kuchimanchi:2012te},  $L_{4L,4R}, L'_{L,R} \rightarrow \eta L_{4R,4L}, \eta L'_{R,L}$ with $\eta = 1$ or $i$ with all other particles having intrinsic parity $1$ as shown in~(\ref{eq:eta}) .  If we choose $\eta = i$ then $P^2 = Z_2$ is an automatic symmetry (as $P \Rightarrow P^2$) that is unbroken by the scalar VEVs.  Therefore the lightest $P^2$ odd heavy lepton is stable and can potentially make up the dark matter. With this choice, the leptonic CP phases will not be generated~\cite{Kuchimanchi:2012te}.

We will now proceed with the choice, $\eta = 1$. 




The leptons have the usual Majorana and Dirac Yukawa couplings given by $f_{ij}(L_{iL}^T\tau_2 \Delta_L L_{jL} + L_{iR}^T\tau_2 \Delta_R L_{jR})$ and $h^\ell_{ij} \bar{L}_{iL} \phi L_{jR}, \ \tilde{h}^\ell_{ij} \bar{L}_{iL} \tilde{\phi} L_{jR} $ where $h^\ell$,  $\tilde{h}^\ell$ are Hermitian due to P. The $i,j$ now run from 1 to 4.  While $f, h^\ell, \tilde{h}^\ell$ are originally real due to CP, they acquire complex phases when we integrate out the heavy leptons (with mass $M^\ell$), just like in the quark sector.  Therefore leptonic CP violation can be generated.   

If $v_R < M^\ell$  the leptonic CP phases generated at the scale $M^\ell$ can help with leptogenesis at the $SU(2)_R \times U(1)_{B-L}$ breaking scale. 

However the challenge is that, if $v_R < M^\ell$, then the resulting complex Yukawa couplings can radiatively generate a strong CP phase in just one loop.  As shown in Reference~\cite{Kuchimanchi_2015} the complex phases in the leptonic Yukawa matrices can provide an imaginary contribution ($\sim Im Tr(ff^\dagger h^\ell \tilde{h}^\ell)$)  to the dimensionless quartic Higgs coupling $\alpha_2$ of the term $\alpha_{2} Tr (\tilde{\phi}^\dagger \phi \Delta_R^\dagger \Delta_R + R \rightarrow L, \phi \rightarrow \phi^\dagger)$ in one loop.  When $\Delta_R$ picks up a VEV, so that both P and CP are now broken,  $\bar{\theta} \sim (m_t/m_b) Im(\alpha_2)$ is generated.

Therefore the leptonic Yukawa couplings must satisfy the severe constraint~\cite{Kuchimanchi_2015} 
\begin{equation}
|Im Tr(f^\dagger f h^\ell \tilde{h}^\ell)| \leq 10^{-11}
\label{eq:lepcpconstraint}
\end{equation}

This implies that CP violation is immeasurably small (or absent) in the leptonic sector, or if it is present some of the Yukawa couplings such as the matrix elements of $f$ must be small, so that most of the parameter space where they are larger is ruled out. Also note that~(\ref{eq:lepcpconstraint})  is almost as constraining as $\bar{\theta} \leq 10^{-10}$, which we are trying to explain.   Therefore generating leptonic CP violation above the scale $v_R$ comes with the above roadblock.

The case $v_R > M^\ell$ has not been discussed before.   Note that in this case the neutrinos in $SU(2)_R$ doublets $L_{iR}$ and $L'_L$ pick up Majorana masses above the soft CP breaking scale $M^\ell$. Therefore there is no CP violation generated in the right handed Majorana neutrino masses. This is not ideal for leptogenesis at this scale.

Below the scale $v_R$ we only have the usual SM particles (without the 3 right handed neutrinos) and a heavy $SU(2)_L$ doublet family composed of $L_{4L}$ with $L'_R$ and a charged iso-singlet vector-like heavy electron. There is now an effective Weinberg type term of the $L_i HHL_j$ with $i,j$ going from 1 to 4 with real CP conserving coefficients.  

Below the scale $M^\ell$ we integrate the heavy leptons by going into their mass basis. This basis rotation leads to CP violating phases appear in the light $3 \times 3$ sector of the Weinberg term (and hence in the light $3 \times 3$ neutrino mass matrix) as well as in the usual charged lepton mass matrices. 

 Though CP violation observable in ongoing neutrino experiments is generated in the PMNS matrix,  this scenario is not entirely satisfactory either as it is not generated in the Majorana masses of the right handed neutrinos. Moreover if for the quarks $v_R < M$ so that the radiatively generated $\bar{\theta}$ is suppressed by the factor $(v_R/M)^2$ (as in equation~\ref{eq:P}), then we'd have a skewed situation with CP violation generated above the P breaking scale for quarks, and below for leptons, which is also not very desirable.



 
  
\section{Soft to Spontaneous CP breaking}
\label{sec:spon}
 
So far we have discussed soft CP breaking.  As noted in Reference~\cite{Kuchimanchi:2010xs},  the introduction of a CP-odd, P-even real scalar singlet $\sigma$ does not spoil the strong CP solution. Therefore we can impose both P and CP as exact symmetries and break both of them spontaneously.  Crucially the term $\mu_\sigma \sigma Tr( \tilde{\phi}^\dagger \phi) + hc$ that can generate the strong CP phase on the breaking of $CP$, is absent as $\mu_\sigma$ must be real due to P and purely imaginary due to CP.

Since CP is imposed, $M_i$ ($i = 1$ to $4$) in equation~\ref{eq:masstermsta} are now all real.  The Yukawa couplings $i h^\sigma_i \sigma (\bar{Q}_{iL}Q'_R + L \leftrightarrow R)$ with $h^\sigma_i$ real due to CP, generate CP violation on mixing with $M_i$, when $\sigma$ picks up a real vev. 

When we imposed the chiral symmetry to obtain the Nelson-Barr form the symmetry also sets $h^\sigma_4$ to zero. While if NB form is not strictly required, then $h_4^\sigma$ can also be present.  

Regardless of the NB form, it appears as if introducing a real scalar and breaking CP spontaneously generates a coincidence of scale problem -- since the scale of the real $M_i$ must be close to the scale of $ih_i^\sigma \left<\sigma\right>$.  However there is no such coincidence of scale issue if we introduce a complex scalar as below.

We now replace the real scalar singlet $\sigma$ with the complex scalar singlet $S$, such that under P, $S \rightarrow S$ and under CP $S \rightarrow S^\star$.  When $S$ picks up a complex VEV spontaneously this breaks CP (while respecting P), and therefore does not generate the strong CP phase.  

This is because terms in the Higgs potential such as $(S^n + S^{\star n}) Tr \tilde{\phi}^\dagger \phi + h.c.$ with real coefficients (due to CP) have the plus sign in the round brackets due to P, where $n$ is a positive integer.  Due to the plus sign no complex phase is generated in terms that involve the bidoublet $\phi$, even after $S$ picks a complex VEV, and therefore we continue to have the strong CP solution.  

The P invariant Yukawa couplings $(h_i S + h'_i S^\star) (\bar{Q}_{iL} Q'_R + \bar{Q}_{iR} Q'_L) + h.c.$ (with $h_i, h'_i$ real due to CP) generate complex mass terms $h_i \left<S\right> + h'_i \left<S\right>^\star$ (that add to the $M_i$ of equation~\ref{eq:masstermsta}).  
In fact we can set $M_i = 0$ (in equation~\ref{eq:masstermsta}) using a $Z_2$ symmetry under which $Q'_R, Q'_L$ and $S$ are odd while all other fields are even. This enables the heavy quarks mass terms to be generated entirely from the VEV of $S$, thereby showing that there is no coincidence of scales problem.  

Note that the $Z_2$ symmetry is anyway automatically present on the dimensionless parameters and would have been softly broken by $M_i$, and in absence of $M_i$ is now spontaneously broken. Non-renormalizable terms with soft and/or spontaneously broken chiral symmetries are discussed in Section~\ref{sec:nr}.

It is also interesting that instead of $Z_2$ we can impose a $Z_4$ symmetry under which $ S \rightarrow i S, Q_{iL}, Q_{iR} \rightarrow iQ_{iL}, i Q_{iR}$ for subscript $i = 1,2,3$ and $Q_{4L}, Q_{4R} \rightarrow -i Q_{4L}, -i Q_{4R}$, so that $S$ only has Yukawa terms such as $S \bar{Q}_{iL} Q'_R$ and $S^\star \bar{Q}_{4L} Q'_R + L \leftrightarrow R, S \rightarrow S$ (where $i = 1,2,3$).  Terms such as $S^\star\bar{Q}_{iL} Q'_R$ (with $i = 1,2,3$) and $S\bar{Q}_{4L} Q'_R$  are absent due to $Z_4$. Moreover $M_i \bar{Q}_{iL} Q'_R$ in equation~(\ref{eq:masstermsta}) are  absent due to $Z_4$ for $i = 1$ to $4$. 

The only term in the scalar potential that depends on the phase of $S$ is the $P, CP$ and $Z_4$ invariant term $\lambda_s (S^4 + S^{\star 4})$ with $\lambda_s$ real due to CP. For $\lambda_s > 0$ this term is minimized when $Arg \left< S \right> = \pm \pi/4$ and therefore the relative phase between $\left<S\right>$ and $\left<S\right>^\star$ which is the relevant phase for CP violation is determined to be $\pm \pi/2$. 

Imposing $Z_4$ also helps reduce the contribution from a non-renormalizable term to $\bar{\theta}$ as we see in Section~\ref{sec:nr}.



 \section{Automatic NB solution with $SU(2)_R$}
\label{sec:mpnb}
We now consider the gauge group $SU(3)_c \times SU(2)_L \times SU(2)_R \times U(1)_X$ with doublet Higgses $H_L$ and $H_R$, that usually occur  with mirror (or generalized) parity~\cite{PhysRevD.41.1286, Craig:2020bnv, Babu:2022ikf}~\footnote{The $U(1)_X$ group has sometimes been given the label of $U(1)_{B-L}$~\cite{Babu:2022jbn}.  As noted in~\cite{Babu:2022ikf} since in mirror parity models the subscript $B-L$ is not actually the difference between the usual Baryon and Lepton numbers for \textit{all} the quarks and leptons, it maybe more appropriate to use the subscript $X$ or $\hat{Y}$ instead.}. The VEV of $H_R$ breaks $SU(2)_R \times U(1)_X \rightarrow U(1)_Y$ with hypercharge $Y = I_{3R} + X/2$, and $Q_{em} = I_{3L} + Y$.  The scalar and fermion content is shown in  Table~\ref{tab:mp}, where we have allowed for any number $m$ of mirror generations, including just one.

Surprisingly we find that there is a strong CP solution without requiring mirror parity, any additional scalars, or needing all 3 mirror generations. As we will now see, once CP is imposed to make Yukawa couplings real and set $\theta_{QCD}=0$, the quark mass matrices have an \itshape{automatic} \normalfont Nelson-Barr form  in this model, and therefore the most general soft-CP breaking dimension 3 mass parameters do not generate $\bar{\theta}$ at the tree-level.  Thus, as we shall see, CP by itself solves the strong CP problem without requiring any other symmetries in this model.

This is interesting as CP itself can be a discrete gauge symmetry~\cite{Choi:1992xp}, and since the Nelson-Barr form is automatic due to the gauged $SU(2)_R$, the smallness of $\bar{\theta}$ can be entirely accidental.

\begin{table}[h]
\centering
\begin{tabular}{cc}
\hline 
& $SU(3)_c \times SU(2)_L \times SU(2)_R \times U(1)_{X}$ \\ \hline \\
$H_R$ & $(1,1,2,1)$ \\ 
$H_L$ & $(1,2,1,1)$ \\  \\ \hline \\
$q_{iL} $ & $(3,2,1,1/3)$ \\ 
$u_{iR} $ & $(3,1,1,4/3)$ \\ 
$d_{iR} $ & $(3,1,1,-2/3)$ \\ 
$l_{iL} $ & $(1,2,1,-1)$ \\ 
$\nu_{iR} $ & $(1,1,1,0)$ \\ 
$e_{iR} $ & $(1,1,1,-2)$ \\ 
$Q_{\alpha R} $ & $(3,1,2,1/3)$ \\ 
$U_{\alpha L}$ & $(3,1,1,4/3)$ \\
$D_{\alpha L}$ & $(3,1,1,-2/3)$ \\
$L_{\alpha R} $ & $(1,1,2,-1)$ \\
$N_{\alpha L}$ & $(1,1,1,0)$ \\
$E_{\alpha L}$ & $(1,1,1,-2)$ \\ \\ \hline
\end{tabular}
\caption{Fermion content of NB solution featuring $SU(2)_R$ with scalars $H_L$ and $H_R$ that are usually associated with mirror parity (or Babu-Mohapatra) solution.  Note that i = 1 to 3 for the 3 usual generations, while the number of mirror generations $m$ need not be the same as the usual generations, as we don't necessarily impose mirror parity.  Therefore $\alpha$ runs from $1$ to $m$ where $m$ is either $1, 2$ or $3.$  Note that the gauge singlet  neutrinos are not required by gauge symmetry, and therefore could be in different numbers though we have taken them to be the same as their corresponding charged fermions. }
\label{tab:mp}
\end{table}

We begin by noting that the Yukawa terms are given by 
\begin{eqnarray}
& & y^u_{ij}  \bar{q}_{iL} H_L u_{jR} + y^d_{ij} \bar{q}_{iL} \tilde{H}_L d_{jR} \nonumber \\ &+& y^U_{{\alpha}\beta}  \bar{U}_{\alpha L} H^\dagger_R Q_{\beta R} + y^D_{{\alpha}\beta} \bar{D}_{\alpha L} \tilde{H}^\dagger_R Q_{\alpha R} 
\label{eq:mpyuk}
\end{eqnarray}
with matrix elements of $y^{u,d}, y^{U,D}$ real due to CP and $i,j$ having values $1$ to $3$,  and $\alpha, \beta$ having values $1$ to $m$  where $m$ can be 1, 2, or 3, depending on the number of mirror families in the model we consider. Note that $\tilde{H}_{L,R} = i \tau_2 H^\star_{L,R}$.

CP is softly broken by 
\begin{equation}
M^U_{{\alpha}j} \bar{U}_{\alpha L} u_{jR} + M^D_{{\alpha}j} \bar{D}_{{\alpha}L} d_{jR}
\label{eq:mpsoft}
\end{equation}

Crucially using an $SU(2)_R$ rotation, the VEV $\left<H^0_R\right> \sim v_R$ of neutral component of $H_R$  can always be chosen to be real and positive, and so when $SU(2)_R \times U(1)_X$ breaks to the SM, the following real CP conserving mass terms are generated from~(\ref{eq:mpyuk})
\begin{equation}
v_R y^U_{{\alpha}\beta} \bar{U}_{\alpha L} U_{{\beta}R} + v_R y^D_{{\alpha}\beta} \bar{D}_{\alpha L} D_{{\beta}R}
\label{eq:real}
\end{equation}
where $U_{\beta R}$ and $D_{\beta R}$ are the up and down components of the $SU(2)_R$ doublet $Q_{\beta R}$ (shown as $Q_{\alpha R}$ in Table~\ref{tab:mp}).

The neutral component of the SM Higgs $H_L$ obtains a VEV $\sim v_{wk}$ that can always be chosen to be real, and  we can see using equations~(\ref{eq:mpyuk}), (\ref{eq:mpsoft}) and (\ref{eq:real}) that the up and down quark mass matrices \textit{automatically} have the Nelson-Barr form:

\begin{equation}
M_{d} =
\left(\begin{array}{cc}      
 y^d v_{wk}  & 0 \\
 M^D  & y^D v_R \\
\end{array}\right)
\label{eq:mpnbform}
\end{equation} 
where all parameters except those in $M^D$ (and $M^U$) are real.  By the usual NB mechanism~\cite{Bento:1991ez} via u-mediation and d-mediation (we have an equal number of heavy $SU(2)_L$ singlet up and down quarks) we can obtain $\delta_{CKM}$ in the light $3 \times 3$ sector, on going to the heavy quark mass basis.   

Radiative corrections for u-mediation and d-mediation using several $SU(2)_L$ singlet heavy quarks have been studied in Reference~\cite{Valenti_2021} and $\bar{\theta}$ is well within experimental bounds for 2 or fewer heavy ups and several heavy downs.  Though either several heavy ups, or several heavy downs, have been considered in these studies,  we have a mixed case of $m$ heavy ups and $m$ heavy downs for $m$ mirror generations.

For $m=1$, we expect the radiative corrections to $\bar{\theta}$  will be well within experimental bounds for the region of  parameter space involving either u-mediation or d-mediation, through just one heavy quark, that generates $\delta_{CKM}$, while the other heavy quark participates trivially.  Even in the over-all parameter space, the correction is expected to be generated in 3-loops and would be small~\footnote{Private communication from Luca Vecchi. The mixed case would be a 3-loop effect similar to the case with either u- or d-mediation, and can be better than u-mediation alone owing to the alternate route of d-mediation.}.  

$m=3$ is a special case where we can also impose mirror parity and obtain the Babu-Mohapatra model~\cite{PhysRevD.41.1286}.  We can replace the subscripts $\alpha$ by $i$ in Table~\ref{tab:mp} and under mirror parity, $q_{iL} \leftrightarrow Q_{iR}, u_{iR} \leftrightarrow U_{iL}, d_{iR} \leftrightarrow D_{iL}$, $H_L \leftrightarrow H_R$ and likewise for the leptons.   This then makes $M^{U,D}$ Hermitian and relates the Yukawa couplings $y^{u,d} = y^{{U,D}^\dagger}$.  $P$ is softly broken by dimension 2 mass parameters of terms $\mu^2_L H^\dagger_L H_L + \mu^2_R H^\dagger_R H_LR$ so that VEVs of $H_L$ and $H_R$ can both be non-zero, while being unequal.

Since the Yukawa couplings in~(\ref{eq:mpyuk}) are real,  $\delta_{CKM}$ is generated by the Nelson-Barr mechanism. Note that we can obtain the usual seesaw form of the Babu-Mohapatra model by interchanging the first and second columns of the NB form of the matrix in equation~(\ref{eq:mpnbform}).

With 3 heavy up quarks and 3 heavy down quarks, the radiative corrections to $\bar{\theta}$ stemming from the NB mechanism, can be interesting (even large) based on the general analysis in Reference~\cite{Valenti_2021}.  Radiative corrections have also been calculated in the Babu-Mohapatra model recently (without the NB mechanism/form) and they are also in general found to be in an interesting~\cite{Hisano:2023izx} or large~\cite{deVries:2021pzl} range.  

Since we now have mirror parity we can allow CP to be broken by all the parameters of the model, as has been usually considered with Babu-Mohapatra model. Also we can go to the other extreme and set  $M^{U,D} =0$ (or small) in equation~(\ref{eq:mpsoft}) by imposing a $Z_2$ symmetry (or approximate symmetry). In this case we can obtain $\delta_{CKM}$ from the complex Yukawa couplings, exactly as in the usual SM. There would be a canceling contribution to the strong CP phase at the tree-level from the mirror sector, and the radiative corrections will likely be negligible, like in the SM. 

For any $m$ with CP imposed, note that we continue to have an \textit{automatic} Nelson-Barr solution even if we break CP spontaneously (instead of softly) by including a scalar singlet $S$ so that the mass terms in~(\ref{eq:mpsoft}) are now real.  Complex phases are generated by the Yukawa couplings of terms such as  $S  \bar{D}_{{\alpha}L} d_{jR}$ and $S^\star  \bar{D}_{{\alpha}L} d_{jR}$ when $S$ picks a VEV, thereby breaking CP spontaneously instead of softly. Note that $S$ can also be a CP odd real scalar singlet.

Since the solution is automatic, there is no `required' symmetry that needs to be imposed to obtain the NB form. Therefore unlike in the last paragraph of section~\ref{subsec:abs} where the required symmetry could be imposed to either allow or prevent leptonic CP violation,  $S \nu^T_R \nu_R$ (and with $S^\star$) are now allowed and produce leptonic CP violation.   Unless they are specifically forbidden by imposing an additional symmetry that isn't required for the NB form.    

Of course if CP is broken softly without the scalar $S$, then whether to also impose soft CP breaking in the leptonic sector is a matter of choice.


\section{SM with soft CP breaking}
\label{sec:minsm}
For completeness, motivated by mirror parity inspired NB solution,  we also consider just the SM with the addition of a vector like singlet heavy quark with components $D_L, D_R$ (both with the same SM gauge quantum numbers as the usual right handed down quarks $d_{iR}$), and a real CP even scalar singlet $\sigma$ with Yukawa term $\sigma \bar{D}_L D_R$.  In this case note that imposing CP on dimensionless parameters (real Yukawa couplings, and $\theta_{QCD} = 0$), and having a symmetry under $D_R \rightarrow -D_R, \sigma \rightarrow -\sigma$ (broken spontaneously by real VEV of $\sigma$) generates the NB form, with CP  broken softly by  dimension 3 mass terms $M^D_i \bar{D}_L d_{iR}$, with complex $M^D_i$.  

This turns out to be a slightly more minimal way of achieving NB solution than the minimal model of References~\cite{Bento:1991ez,alves2023vectorlike}.

\section{Non-renormalizable terms}
\label{sec:nr}

We will consider non-renormalizable terms suppressed by high energy scale $\Lambda$ (or the Plank scale). We will begin with the particle content of Table~\ref{tab:matterta}, that has the triplet and bi-doublet Higgses,  and includes a heavy quark family.  Regardless of the strong CP phase, note that the term
\begin{eqnarray}
& & \bar{Q}_{iL} Q'_R Tr ( \Delta_R^\dagger \Delta_R)/\Lambda  + \bar{Q}_{iR} Q'_L Tr ( \Delta_L^\dagger \Delta_L)/\Lambda \nonumber \\ & &\sim (v^2_R/ \Lambda) \bar{Q}_{iL} Q'_R 
\label{eq:nonherm}
\end{eqnarray}

contributes a mass $\sim v_R^2/\Lambda \sim 10^8 GeV$ to  the heavy quark mass, where we have taken for example, $v_R \sim 10^{13} GeV$ and $\Lambda \sim 10^{18} GeV$. Therefore if the coefficient of the above term is O(1), then it would appear as if heavy fermions, whose masses $\sim M$ are protected by chiral symmetries (such as under $Q'_R \rightarrow e^{i\alpha}Q'_R$) and therefore can be naturally light, would require fine-tuning if they are lighter than $10^8$ GeV. 

Therefore we suppress the above term by a factor $M/\Lambda$, so that it is protected by the same chiral symmetry, and vanishes as $M \rightarrow 0$ and the symmetry is restored. In other words we consider non-renormalizable terms that, like renormalizable terms,  are protected by approximate (or softly broken) symmetries.

With this ansatz the above term can be re-written as    
\begin{equation}
c_i (M/\Lambda) \bar{Q}_{iL} Q'_R Tr ( \Delta_R^\dagger \Delta_R)/\Lambda \sim c_i M (v_R/ \Lambda)^2 \bar{Q}_{iL} Q'_R 
\label{eq:alt}
\end{equation}
where $c_i$ are complex. Since the above term is not Hermitian (as $\left<\Delta_L\right> << v_R$ in equation~(\ref{eq:nonherm})), it generates 
\begin{equation}
\bar{\theta}_{Non-Renorm} \sim c_i (M/\Lambda)(v_R^2/\Lambda)/M \sim  c_i (v_R/ \Lambda)^2 
\label{eq:thetanon}
\end{equation}
when it is considered along with the Hermitian terms of equation~(\ref{eq:masstermsta}). This is acceptably small for $v_R \leq 10^{13} Gev$, which just about includes the canonical seesaw scale and $\Lambda \sim 10^{18} GeV$. 

It is interesting that this contribution to $\bar{\theta}$ does not depend on $M$, or even whether $M$ is larger or smaller than $v_R$.  Nor does it vanish for the Nelson-Barr form in equation~(\ref{eq:masstermsta}) since $c_4$ can be complex. 

Instead of dealing with the soft-breaking of chiral symmetry in the above manner,  we can introduce the complex scalar singlet $S$ of Section~\ref{sec:spon} and consider non-renormalizable terms that respect $P, CP$ and spontaneously broken $Z_2$ under which $Q'_L, Q'_R$ and $S$ are odd and other fields are even. 

Due to $Z_2$ the above term in equation~(\ref{eq:nonherm}) and~(\ref{eq:alt})  is absent, and instead we have $Z_2$ invariant terms such as 
\begin{equation}
r_i S \bar{Q}_{iL} Q'_R Tr ( \Delta_R^\dagger \Delta_R)/\Lambda^2 + L \leftrightarrow R
\end{equation}
 with real couplings $r_i$,  and $|\left<S\right>|$ sets the mass scale of the heavy quarks. Since $\left<S\right>$ is complex and violates $CP$,  $\bar{\theta} \sim (r_i \left<S\right>/ |\left<S\right>|) (v_R/\Lambda)^2 \sim (v_R/\Lambda)^2 $ is induced when P is broken at the scale $v_R$,  which is consistent with the previous result (equation~(\ref{eq:thetanon})) from soft breaking of chiral symmetry.  
, 

Contribution of $\bar{\theta} \sim (\left<S\right>/\Lambda)^2$  can be found by considering the $Z_2$ invariant non-renormalizable term $S^2 Tr (\tilde{\phi}^\dagger \phi \Delta^\dagger_R \Delta_R)/\Lambda^2$, with the heavy quark mass scale set by $|\left<S\right>|$. This is because the vev $\kappa_2$ of the bi-doublet $\phi$ picks up a complex phase $\sim (\left<S\right>/\Lambda)^2$. However, this $Z_2$ invariant term is absent and the contribution is further suppressed so that $\bar{\theta} \sim (\left<S \right>/\Lambda)^4$  if the $Z_4$ symmetry discussed at the end of Section~\ref{sec:spon} is introduced.  In this case the $Z_4$ symmetric non-renormalizable term is $S^4 Tr (\tilde{\phi}^\dagger \phi \Delta^\dagger_R \Delta_R)/\Lambda^4$. Such a non-renormalizable term is absent in the minimal model of Table~\ref{tab:matterta} without the singlet $S$.  



While we have discussed generic non-renormalizable terms, we also note that without specific knowledge of these terms and a consistent way of evaluating their loop corrections, it may be better just to focus on the renormalizable terms and the testable predictions, notably the absence of leptonic CP violation for the minimal model of Table~\ref{tab:matterta}.

Now for the mirror-parity inspired particle content of Table~\ref{tab:mp}, with doublet Higgses (rather than the bi-doublet), and having the automatic Nelson Barr solution discussed in section~\ref{sec:mpnb}, the zero in the matrix~(\ref{eq:mpnbform}) will get a correction due to the non-renormalizable term $\bar{q}_{iL} H_L H^\dagger_R Q_{\alpha R}/\Lambda$ which generates (on evaluation of the determinant)  $\bar{\theta} \sim M^D/\Lambda \sim v_R/\Lambda$  for the case $m=1$, that is with one mirror family and no mirror parity. Thus we would expect the heavy mirror quarks to be at a scale $v_R \leq 10^{8} GeV$ for $\bar{\theta} \leq 10^{-10}$, for the particle content of Table~\ref{tab:mp}.

\section{Conclusions}
\label{sec:conc}

While P and CP have historically been treated as two different approaches to the strong CP problem, in this work we find that in the popular left-right symmetric model based on $SU(2)_L \times SU(2)_R \times U(1)_{B-L}$ with bi-doublet Higgs, where P sets $\theta_{QCD}$ to zero, the heavy quark family needed to generate the CKM phase can have couplings of the Nelson-Barr form which has so far only been seen in solutions with spontaneous CP violation. The NB solution we obtain  does not have the problem of requiring a close coincidence of scales between CP breaking VEVs and CP conserving mass terms, which is the vexing issue that all other NB solutions have.  Moreover P protects $\bar{\theta}$ if it breaks at a scale $v_R$ below the mass $M$ of the heavy quarks.  Thus in our model $\bar{\theta} \sim 10^{-9} (v_R/M)^2$ is generated which can be sufficiently small even if $M$ is only an order of magnitude larger than $v_R$.   

If the heavy masses  are below the P breaking scale, so that $M < v_R$,  to our surprise we find irreducible corrections under RGE running to $\bar{\theta} \sim (10^{-13}~to~10^{-10}) ln(v_R/M)$ that are in the reach of ongoing neutron EDM experiments. If experiments discover a neutron EDM consistent with $\bar{\theta} \geq 10^{-12}$ then it may be worth looking for these heavy quarks that have very small Yukawa couplings with the usual quarks in future colliders.    $\delta_{CKM}$ in this case is not generated through the NB mechanism, as the latter gives a much higher irreducible $\bar{\theta} \sim 10^{-9}$. 

An exciting testable prediction is the absence of leptonic CP violation in the minimal model we consider in Table~\ref{tab:matterta}.  We predict $sin(\delta_{CP}) = 0$ for the Dirac phase of the PMNS matrix. Global fits to current neutrino experiments' data are consistent with this prediction and we look forward to future experiments with greater sensitivity. The absence of leptonic $\delta_{CP}$ has been discussed before~\cite{Kuchimanchi:2012xb,Kuchimanchi:2012te} in the context of Reference~\cite{Kuchimanchi:2010xs}, and in ~\cite{Kuchimanchi_2015}.  We also discuss some challenges in a non-minimal model that can generate leptonic CP violation.


Last but not the least, we find that in models with mirror (or generalized) parity, and containing doublet rather than the bi-doublet Higgses, there is an automatic Nelson-Barr solution on imposing CP. This is a new and  more economical approach to addressing the strong CP problem with $SU(2)_L \times SU(2)_R $ group and respective doublet Higgses (rather than the bi-doublet), since we do not have to impose mirror parity and can also have just 1 mirror generation.              

It is interesting that mirror parity inspired us to find a solution to the strong CP puzzle where CP by itself solves the strong CP problem, and P is not imposed, while the NB form is accidental due to the gauged $SU(2)_R$.  If CP is a discrete gauge symmetry and the Nelson-Barr form is accidental due to a gauged $SU(2)_R$, then the smallness of $\bar{\theta}$ can be entirely accidental.   

\section*{Acknowledgment}  
I thank Luca Vecchi for the correspondence on radiative contribution discussed in Section~\ref{sec:mpnb}.
 
\raggedright 
\bibliography{nelson}


\end{document}